\documentclass[aps,prb,draft,groupedaddress]{revtex4}
\usepackage{graphicx}
\begin{document}
\title{Remarkable thermal stability of BF$_3$-doped polyaniline}
\author{Debangshu Chaudhuri}
\author{Prashanth W. Menezes}
\author{D. D. Sarma$^{a)}$}
\affiliation{Solid State and Structural Chemistry Unit, Indian
Institute of Science, Bangalore-560012, India}

\begin{abstract} We show that the recently synthesized BF$_3$-doped polyaniline (PANI)
exhibits remarkable stability against thermal ageing. Unlike the
protonated PANI, which shows rapid degradation of the conductivity
on heating in air, BF$_3$-doped PANI shows more than an order of
magnitude improvement in conductivity. We employ x-ray
photoelectron spectroscopy (XPS), fourier transform infra-red
(FTIR) spectroscopy, and x-ray diffraction (XRD) to
understand this unexpected phenomenon.

\end{abstract}
\maketitle

Interesting electrical and optical properties exhibited by
conducting polymers make
them an attractive choice for various technological applications.
However, the use is often limited by a number of crucial technical factors
such as the environmental stability. Rapid degradation on exposure to
environmental conditions is the primary reason why many of the doped
polyacetylene systems, with conductivities nearly as high as
that of copper, find less use in applications than doped polyaniline (PANI)
with much lower conductivities.~\cite{1,2} The 50\% oxidized form of
PANI is popular for its environmental stability in both
pristine and doped forms. However, at elevated temperatures (above~$\sim$75~$^\circ$C)
the doped PANI suffers significant drop in the conductivity.~\cite{3,4} The
process of thermal ageing in HCl-doped PANI has been studied extensively
and contrary to the previously held hypothesis of thermal dedoping, it was
found that the polymer undergoes a chain oxidation followed by chlorination
of the aromatic rings, causing a serious disruption in the conjugated network.~\cite{3,5}
Since the protonated polymer undergoes irreversible chemical modifications
upon heating in air, its conductivity cannot be recovered by redoping.~\cite{6}
In several respects, the
recently discovered form of highly conducting PANI doped with electron-deficient
BF$_3$ is distinct compared to the usual protonated form of
PANI.~\cite{7,8} In particular, a high conductivity in conjunction
with its amorphous nature makes this new system quite interesting.~\cite{8}
However, nothing is as yet known of the environmental stability of this interesting
class of compounds. In the present work, we report a detailed investigation of various
thermal treatments of BF$_3$-doped PANI in order to understand the effect of
thermal ageing. We find a remarkable thermal stability of BF$_3$-doped PANI compared
to other doped PANI systems, the thermal treatment in fact leading to an unexpected
increase in the conductivity over a wide range of temperature and duration of the
treatment.

The sample was prepared~\cite{7} by doping BF$_3$ into the
emeraldine base, synthesized by the chemical oxidation of aniline
using (NH$_4$)$_2$S$_2$O$_8$ at 0-5~$^\circ$C.~\cite{9} In order to investigate
thermal ageing in normal atmospheric conditions, powder samples were heated at
two temperatures, 100 and 140~$^\circ$C, for different time intervals in an open
mouthed vessel; for comparison, heating experiments were also performed in an
argon atmosphere. The powder was characterized at different stages of heating using
fourier transform infra-red (FTIR) spectroscopy and x-ray photoelectron
spectroscopy (XPS). Resistivity was measured by pressing the powder into thin pellets.
In certain cases, pellets instead of powders were given the thermal treatment.

Fig.~\ref{air}\emph{a} shows the effect of heating powdered BF$_3$-doped and
HCl-doped PANI in air at 100 and 140~$^\circ$C; we also show the results of heating
pellets at 100~$^\circ$C in the same figure. The relative
conductivity, defined as the ratio of conductivity of the aged
sample to that of the non-aged one, is plotted against the duration of heating.
HCl-doped PANI clearly shows a marked degradation upon heating.
The conductivity of the powdered samples decreases by orders of
magnitude, the higher temperature heating exhibiting a much
stronger effect; even the pellet shows a substantial loss in
conductivity at 100~$^\circ$C. On the contrary, BF$_3$-doped PANI
shows an \emph{increase} of the conductivity in every case, particularly
in the short time regime, in certain cases the samples showing
more than an order of magnitude improvement. Additionally, beyond
the initial stages of heating, the conductivity remains almost the
same in spite of prolonged heating in every case.

Various experiments carried out by us suggest that heating in
air or in Ar yields very similar results, eliminating the
possibility of any oxidation related influence on the
conductivity. Fig.~\ref{air}\emph{b} shows the change in relative
conductivity upon heating the pellet in Ar at various temperatures
for two different time intervals. We note that the conductivity
reaches its highest value upon heating at temperatures close to
125~$^\circ$C. In fact, among all the heated samples, the highest
conductivity (34.2~S~cm$^{-1}$) was obtained upon heating the
pellet at 125~$^\circ$C for 1 hr. Further, we observed that
annealing the sample at 125~$^\circ$C for a period as short as 2
min. is sufficient to increase the conductivity by an order of
magnitude. The experimental data presented in Fig.~\ref{air}\emph{a} and
\emph{b} are sufficient to establish the remarkable thermal
stability of BF$_3$-doped PANI in sharp contrast to protonated
PANI and consequently the desirability and efficacy of this
recently discovered system, particularly in high temperature
applications. In the following, we probe the microscopic origin of
this unexpected behavior.

We find that the temperature-dependence of conductivity in all
BF$_3$-doped PANI samples is well described in terms of Sheng's
model~\cite{10,11} of fluctuation induced tunnelling, leading to
$\sigma(T)=\sigma_0~exp~[-T_0/(T+T_1)]$. Fig.~\ref{sheng} shows
the excellent linearity of $ln~\sigma$ for a number of samples
with a variety of treatment, when plotted against $1/(T+T_1)$,
where $T_1$ in each case is determined by a least-squared error
fitting procedure. The parameter values obtained from the fit are
listed in Table~\ref{fit}. Sheng's model views the doped polymer
as a heterogeneous system, consisting of large, highly conducting
regions separated by thin insulating barriers. In BF$_3$-doped
PANI, these insulating barriers most probably arise from the grain
boundaries. $T_0$ in the above expression for $\sigma(T)$ defines
the average height of the potential barrier between two conducting
grains. Table~\ref{fit} shows that $T_0$ and consequently the
barrier height decrease significantly and systematically upon
heating, suggesting a better intergrain connectivity in the heated
samples, providing a clue to the understanding of the unexpected improvement of
$\sigma(T)$ on heating. Similar evidence is also provided by the x-ray diffraction
(XRD) data.

In the case of HCl-doped PANI, it is known that the ageing process is initiated
at the periphery of the crystalline grains and moves progressively towards its center,
transforming the material into an amorphous and poorly conducting phase.~\cite{3}
As prepared BF$_3$-doped PANI, however, is known to be already
amorphous, despite its high conductivity.~\cite{8} Powder XRD
patterns of the heated samples showed complete resemblance with the
non-heated sample, suggesting their amorphous nature. It was earlier reported~\cite{8}
that the XRD pattern of as-prepared BF$_3$-doped PANI, when left in atmosphere for
sometime, develops a peak at 24.8$^\circ$, which corresponds to the formation of
protonated PANI, as a result of BF$_3$ hydrolysis; simultaneously, two sharp peaks
corresponding to boric acid begin to appear at 14.8$^\circ$ and 28.1$^\circ$.
This suggests that some amount of unreacted BF$_3$ remains adhered to the as-prepared
sample, probably at the grain boundaries, that reacts with moisture in the atmosphere.
In contrast, the XRD patterns of the heated samples do not show the above
mentioned changes even after being left in atmospheric conditions for a long time.
It appears that by heating the sample, it is possible to get rid of this impurity
from the grain boundary surfaces and therefore, it must be at least partially
responsible for the increase in the conductivity.

The macroscopic conductivity of a multigrain sample is contributed by both the
intragrain intrinsic conductivity and the intergrain connectivity. The evidences
presented so far establish only an improved intergrain connectivity in the heated samples.
In order to investigate the possibility of changes in the intragrain chemical composition,
we carried out a detailed XPS study on BF$_3$-doped PANI. Fig.~\ref{xps}\emph{a}
shows the N~1\emph{s} spectra of the doped sample before (solid circles) and after (open
circles) heating at 100~$^\circ$C.
For the ease of comparison, we have overlapped the spectrum of the unheated sample
(solid line) with that of the heated one. It is evident that the two spectra are essentially
identical, establishing identical chemical states for N in both the cases. C~1\emph{s}
spectra, shown in Fig~\ref{xps}\emph{b}, also exhibit identical features for
as-prepared and heated samples, once again demonstrated by overlapping the signal from the
pristine sample as a solid line on the signal from the heated sample (open circles). This
suggests that the thermal treatment does not
change the backbone of the doped polymer. However, B~1\emph{s} spectrum exhibits a
significant decrease in the spectral width upon heating (see Fig.~\ref{xps}\emph{c}).
In order to understand better the spectral changes, we decompose the total B~1\emph{s}
spectrum in each case in terms of component spectra appearing at 193.5$\pm$0.05~eV and
192.4$\pm$0.1~eV shown by thin dashed lines, arising from BF$_3$ reacted to the PANI
and traces of boric acid due to partial hydrolysis of unreacted BF$_3$ near the grain
boundary, respectively; the thick lines overlapping the data
points are the sums of the two components. The spectral decomposition of B~1\emph{s}
spectra clearly establish a strong reduction of the peak at 192.4~eV on heating the sample,
leaving the other B~1\emph{s} component arising from the doped BF$_3$ essentially
unchanged. Since the features corresponding to N~1\emph{s}, C~1\emph{s}
and B~1\emph{s} from the doped BF$_3$ species remain identical in as-prepared and heated
samples, we conclude that heating does not change the chemical composition of the doped
polymer; therefore, the changes in the conductivity on heating are primarily due to changes
near the grain boundaries, rather than any change within the grains.

We have also characterized the samples at every stage of heating
using FTIR spectroscopy. When the sample is heated in air at
140~$^\circ$C for prolonged period of time, the spectrum exhibits a few small changes.
A small peak begins to appear at 1374 cm$^{-1}$ after about 8~hrs of heating, growing
further in intensity with continued heating. Although the origin of this peak remains
unclear, it is known that this peak is present in the spectrum of undoped PANI and
that it diminishes gradually with an increase in the extent of doping,~\cite{7} being
present in all partially doped PANI systems. We, therefore, conclude that heating
for prolonged periods at a high temperature leads to a slight amount of
dedoping; this is consistent with the decrease in the
relative conductivity below unity, as observed in the case of
BF$_3$-doped PANI heated at 140~$^\circ$C for 1500~min. (see
Fig.~\ref{air}\emph{a}). The fact that no such change is observed
in the FTIR spectrum of samples heated at 100~$^\circ$C, is
in agreement with the conductivity results, showing a saturation and suggesting
the absence of any significant dedoping at 100~$^\circ$C. Since, dedoping is
found to be the main reason for the
decrease in the conductivity on prolonged heating at the higher
temperature, we redoped the sample previously heated at 140~$^\circ$C
for 1500~min. The redoped sample showed a 2.6 fold increase in the
room temperature conductivity, leading to a value that is close to the
highest conductivity observed in the case of heating at
140~$^\circ$C.

In conclusion, BF$_3$-doped PANI shows remarkable stability towards thermal ageing.
In sharp contrast to protonated PANI, where heating in air causes chemical degradation and
sharp decrease in the conductivity, BF$_3$-doped PANI shows a rapid enhancement in
conductivity. XPS, FTIR and XRD results show that
heating removes the impurities from the grain boundary regions of the as-prepared
samples. Partial dedoping upon heating at 140~$^\circ$C for long
periods causes a small decrease in the conductivity, which
could however be recovered upon redoping the heated samples.

We are grateful to the Council for Scientific and Industrial Research (CSIR) for
providing the financial support.

%\newpage

%\clearpage
\begin{figure}[h]
\caption{\label{air} (a)~:~Variations of the relative conductivities of HCl and
BF$_3$-doped PANI with the duration of heating. Circles and triangles represent heating of
powder at 100~$^\circ$C and 140~$^\circ$C, respectively. Diamonds represent heating of
pellets at 100~$^\circ$C; open and closed symbols correspond to BF$_3$ and HCl-doped
PANI, respectively. The data for HCl-doped PANI are taken from Ref. 4.
(b)~:~Variations of relative conductivities of BF$_3$-doped PANI pellets for two
different durations of heating in dry Ar as a function of the temperature of heating.}
\end{figure}

\begin{figure}[h]
\caption{\label{sheng} $ln~\sigma$ plotted against $(T+T_1)^{-1}$, where $T_1$ in each case
is obtained from a least-squared error fitting procedure for $\sigma(T)$ according to
Sheng's model (see text).}
\end{figure}

\begin{figure}[h]
\caption{\label{xps} XPS core-level spectra from BF$_3$-doped PANI,
before and after heating:~(\emph{a})~N~1\emph{s}, (\emph{b})~C~1\emph{s} and
(\emph{c})~B~1\emph{s}. Thin solid lines in panels (a) and (b) represent the spectra
of the unheated sample, shifted and overlapped on the spectra (open circles) of the heated
samples for comparison. Thin dashed lines in (c) indicate the two boron components obtained
from a least-squared-error fit, while the solid lines overlapping the data points are
the sum of the two components. Upon heating the boric acid component decreases from
20$\%$ to 5$\%$.}

\end{figure}

%\clearpage
\begin{table}

\caption{\label{fit}Parameters for the conductivity model (see text). A:~before heating,
B:~powder at 140~$^\circ$C in air for
3~hrs., C:~powder at 100~$^\circ$C in air for 3~hrs., D:~pellet at 125~$^\circ$C in Ar for
1~hr., E:~pellet at 100~$^\circ$C in air for 8 hrs.}

\vspace*{0.5cm} \normalsize \centering

\begin{tabular}{c|ccccc}
\hline
  &~~A~~&~~B~~&~~C~~&~~D~~&~~E\\
\hline
$T_0$~(K)&~~1515~~&~~1221~~&~~1013~~&~~644~~&~~578\\
$T_1$~(K)&~~113.4~~&~~60.9~~&~~64.6~~&~~122.4~~&~~102.5\\
$\sigma_0$~(S~cm$^{-1}$)&~~75.4~~&~~97.5~~&~~98.5~~&~~122.0~~&~~130.8\\
\hline
\end{tabular}
\end{table}
\end{document}